\documentclass[amsmath,amssymb,aps,prl,twocolumn, nofootinbib]{revtex4-1}
\usepackage{graphicx, hyperref, csquotes}

\def\extd{\mathrm {d}}
\newcommand{\Tr}{\mathrm{Tr}}
\newcommand{\rk}{r}
\renewcommand{\ell}{c}
\newcommand{\efd}{{d_{\textrm{eff}}}}
\newcommand{\gf}{\phi}
\newcommand{\gfb}{\bar\phi}
\newcommand{\vg}{\pmb{g}}

\newcommand{\mop}{(\gfb\cdot_{\hat{\ell}}\gf)}
\newcommand{\cn}[1]{\lambda_{#1}}
\newcommand{\cnr}[1]{\tilde{\lambda}_{#1}}
\newcommand{\mr}{\tilde{\mu}_k}
\newcommand{\cns}[1]{\bar{\lambda}_{#1}} 
\newcommand{\ms}{\bar{\mu}_k}
\newcommand{\gfc}{W}
\newcommand{\reg}{{\mathcal R}_k}
\newcommand{\wfr}{Z_k}
\newcommand{\gfk}{\Gamma_k}
\newcommand{\ks}{{2}}
\newcommand{\cs}{a}
\newcommand{\ak}{N_k}
\newcommand{\cct}{1}
\newcommand{\ccu}{2}
\newcommand{\vol}[1]{v_{#1}}
\newcommand{\bco}[1]{\beta^{#1}}
\newcommand{\bcb}[1]{\bar{\beta}^{#1}}
\newcommand{\buv}[1]{{\beta}_{\textsc{uv}}^{#1}}
\newcommand{\nmax}{n_\textrm{max}}

\begin{document}

\title{(No) phase transition in tensorial group field theory}

\author{Andreas G. A. Pithis}
\email{pithis@thphys.uni-heidelberg.de}
\affiliation{Scuola  Internazionale  Superiore  di  Studi  Avanzati  (SISSA),\\ via  Bonomea  265,  34136  Trieste,  Italy, EU\\
Institut f\"ur Theoretische Physik, Universit\"at Heidelberg,\\ Philosophenweg 16, 69120 Heidelberg, Germany, EU}

\author{Johannes Th\"urigen}
\email{johannes.thuerigen@uni-muenster.de}
\affiliation{Mathematisches Institut der Westf\"alischen Wilhelms-Universit\"at M\"unster\\ Einsteinstr. 62, 48149 M\"unster, Germany, EU\\
Institut f\"ur Physik/Institut f\"ur Mathematik der Humboldt-Universit\"at zu Berlin\\
Unter den Linden 6, 10099 Berlin, Germany, EU
}

\begin{abstract}
Continuum spacetime is expected to emerge via phase transition in discrete approaches to quantum gravity. A promising example is tensorial group field theory but its phase diagram remains an open issue. The results of recent attempts in terms of the functional renormalization group method remain inconclusive since they are restricted to truncations of low order. We overcome this barrier with a local-potential approximation for $\textrm{U}(1)$ tensor fields at arbitrary rank $\rk$ focusing on a specific class of so-called cyclic-melonic interactions. Projecting onto constant field configurations, we obtain the full set of renormalization-group flow equations. At large cut-offs we find equivalence with $\rk-1$ dimensional $\textrm{O}(N)$ scalar field theory in the large-$N$ limit, modified by a tensor-specific, relatively large anomalous dimension.
However, on small length scales there is equivalence with the corresponding scalar field theory with vanishing dimension and, thus, no phase transition. This is confirmed by numerical analysis of the full non-autonomous equations where we always find symmetry restoration. The essential reason for this effect are isolated zero modes. This result should therefore be true for tensor field theories on any compact domain and including any tensor-invariant interactions. Thus, group field theories with non-compact degrees of freedom will be necessary to describe a phase transition to continuum spacetime.
\end{abstract}


\maketitle

\section{Introduction}

In various approaches to a quantum theory of gravity, continuum spacetime is expected to emerge from a microscopic theory of discrete geometries in terms of a phase transition~\cite{CDT,TMrev,GurauBook,GFTCC1}.
A promising example for such a quantum theory of fundamental geometric degrees of freedom is tensorial group field theory (TGFT)~\cite{TFT, GFT}.
Like matrix models relate to two-dimensional gravity~\cite{MM}, such a theory with tensor fields of rank $\rk$ generates discrete $\rk$-dimensional geometries which converge to continuum geometries at criticality.
It is thus crucial to understand the theory's phase structure.

The renormalization group is a useful tool to understand how physical theories evolve along scales and allows to chart their phase diagrams.
In this way, important first insights have been gained for TGFT using the method of the functional renormalization group (FRG)~\cite{WetterichBergesTetradis,Delamotte, FRGreview, Kopietz}, first applied to discrete geometry systems in Ref.~\cite{EiKos2013}. First of all, the FRG equations are non-autonomous~\cite{BGBeOr2015}.
This is a general consequence of an external scale $\cs$ in a theory. 
In TGFT, it comes from isolated zero modes in the spectrum on a compact domain of size $\cs$.
The tensorial structure of interactions leads to various combinations of zero-modes, and thus a special kind of non-autonomous equations~\cite{BGBeOr2015, TGFTnoncompact,TGFT1, TGFT2, LahocheReview}.

Various insights in the phase structure of TGFT have been gained which are based on truncations of the theory space to low order~\cite{BGBeOr2015,TGFT1,TGFT2,TGFTnoncompact,LahocheReview, BGKoPeOr2018, EiLuKoPe2019, EiLuPeSi2020} or apply to the autonomous UV limit~\cite{TGFT3}. In these works one typically finds non-Gaussian fixed points but due to the named limitations these results need further verification. 
In addition, it has remained elusive so far to study the IR properties of such theories.
In particular, the question has not been settled whether phase transitions and different phases can truly exist for such models. 
 
Here we solve these puzzles establishing a local-potential approximation (LPA$'$)%
\footnote{In the usual FRG jargon, our approximation is an LPA$'$ meaning that we consider also the flow of the anomalous dimension; it is, however, not the full LPA$'$ of TGFT because of the restriction to a specific class of infinitely many interactions.}
restricting to cyclic-melonic interactions at any order. 
Exploiting a projection to constant field configurations, we are for the first time able to explicitly derive the non-autonomous equations for that potential
at arbitrary rank, valid at all scales. 
In the resulting flow equations for couplings at any order of interaction the non-autonomous part factorizes and can be resummed.
Correspondence with the flow equations in $\textrm{O}(N)$ scalar field theories on Euclidean space~\cite{WetterichBergesTetradis,ONMerminWagner} allows then for precise results in the large- and small-scale limits, only modified through a tensor-specific flow of the anomalous dimension.

As a result we find that there is a non-Gaussian fixed point only in the large cut-off limit, and only for $\rk=4$. 
The scaling exponents at this point are altered by the relatively large anomalous dimension.
Flowing to smaller scales this fixed point does not persist and we universally find restoration of the global $\textrm{U}(1)$ symmetry. 
Thus, there is only the unbroken phase in our approximation, that is no phase transitions.
This is in accordance with results on scalar field theory on compact spaces~\cite{Benedetti,Serreau} and with our earlier work on mean-field approximation in group field theory~\cite{Pithis:2018bw}.
We argue that the isolated zero mode due to the compactness of the field domain is the essential reason for symmetry restoration and conjecture that this phenomenon applies to any compact domain and the full theory space including arbitrary tensor-invariant interactions.
 
\section{Functional renormalization group in the melonic potential approximation}
 
In this work we consider TGFT on the group $G=\textrm{U}(1)$ at any rank $\rk\ge3$. 
That is, the field is a complex scalar $\gf:\textrm{U}(1)^\rk\rightarrow \mathbb{C}$.
The field theory is defined by generating functions $Z$ or $\gfc$ 
in terms of an action $S$ and a measure on the space of field configurations $D\gf D\gfb$,
\begin{equation}
Z[J,\bar{J}]\equiv e^{\gfc[J,\bar{J}]} = \int D \gf D \gfb \, e^{-S[\gf,\gfb] + (J,\gf) + (\gf,J)}.
\end{equation}
where $(\gf,\psi)=\int \extd \vg\, \gfb(\vg)\psi(\vg)$ is the scalar product on $L^2(G^\rk)$ using a dimensionful Haar measure $\int\extd\vg = \cs^\rk$.
The volume scale $\cs$ will be necessary for physically meaningful rescaling of quantities. 
Furthermore, it allows to consider the large-volume limit
$\cs\rightarrow\infty$ of the compact manifold corresponding to the theory on $\mathbb{R}^\rk$ where an IR regularization can be removed  exactly by this limit~\cite{TGFTnoncompact}.

\renewcommand{\gf}{\varphi}       
\renewcommand{\gfb}{\bar\varphi}
For the functional renormalization group the natural object to consider is the effective action, that is the Legendre transform exchanging sources for the average field
$\gf(\vg) :=  \langle\phi(\vg)\rangle  =  \frac{\delta W[\bar{J},J]}{\delta \bar{J}(\vg)}$
and respectively
$\gfb(\vg)$. 
Adding an IR regulator $\reg$ depending on the running scale $k$, the effective average action \`{a} la Wetterich is
\begin{equation}
\gfk[\gf,\gfb] = \sup_{\bar{J},J}\{(\gf,J) + (J,\gf) - \gfc_k[\bar{J},J]\} - (\gf, \reg \gf)
\end{equation}
and one can deduce the FRG equation~\cite{Wetterich, Morris, TGFT1}
\begin{equation}\label{eq:Wetterich}
k \partial_k \gfk = \frac{1}{2} \overline{\Tr} \left[ \left( \Gamma_k^{(2)} + \reg \mathbb{I}_\ccu \right)^{-1} k \partial_k \reg \mathbb{I}_\ccu \right]\,,
\end{equation}
with initial conditions $\Gamma_{k=\Lambda}[\gf,\gfb]=S[\gf,\gfb]$ at the UV scale $\Lambda$. Here $\gfk^{(2)}$ is the $2\times2$
Hessian matrix of $\gfk$ with respect to $\gf, \gfb$ and the trace $\overline{\Tr}$ sums over all field degrees of freedom.
    
In this work we restrict the theory space to cyclic-melonic interactions.
It is defined by the effective average action
\begin{equation}\label{eq:model}
\Gamma_k [\gfb , \gf] = \left( \gf, \left( -Z_k  \Delta + \mu_k  \right) \gf\right)
+ \sum_{\ell=1}^\rk \Tr_G V_k\mop
\end{equation}
where $-\Delta = \cs^{-2} C$ is the dimensionful Laplacian on $G^\rk$ in terms of the Casimir $C$, $\wfr$ is the wave-function renormalization parameter and $m_k$ is the coupling at quadratic order.
The potential is determined by the power series
\begin{equation}
V_k(z) = \sum_{n = 2}^{\infty} \frac{1}{n!\,\rk}\cn{n} z^n  
\end{equation}
with $k$-dependent couplings $\cn{n}$ for $n\ge2$.

Melonic interactions are those recursively constructed by inserting a melonic operator $\mop$ of colour $\ell$, defined via its kernel
\begin{equation}
\mop(g_\ell,h_\ell)
:= \int \prod_{b\ne\ell}\extd g_b \gfb(g_1...g_\ell...g_\rk)\gf(g_1...h_\ell...g_\rk ) \, ,
\end{equation}
into one of different colour (in the integral), or concatenating it with one of the same colour~\cite{GurauBook}. 
This construction can be mapped to rooted $\rk$-coloured trees~\cite{BoGuRiRi2011}.
The cyclic-melonic interaction of colour $\ell$ in Eq.~\eqref{eq:model} are those build only from concatenations.
This corresponds to non-branching trees 
in the tree description.

This approximation is less restrictive as it may seem at first sight.
It is true that already the quartic melonic interaction generates any tensor invariant since any bipartite edge-coloured graph occurs as the boundary of a Feynman diagram with such quartic interactions~\cite{PerezSanchez:2018fn}. 
Thus, the truncation to only cyclic-melonic interactions ignores most interactions in the class of possible interactions proliferating with the number of fields.
But 
the cyclic-melonic interactions are known to dominate the flow and provide the relevant directions in the UV~\cite{TGFT3, BGKoPeOr2018, EiLuKoPe2019}. 
More importantly, this restriction allows us in the first place to calculate the FRG flow through all scales and to arbitrary order in the potential. 
We will then see that the qualitative result of symmetry restoration is independent of the details of the specific interactions and should extend to the full tensorial potential.

As common in the FRG method, we project to a constant field configuration in direct space in the FRG equation.
That is, after evaluating the second derivative $\gfk^{(2)}$ we set $\gf(\vg) = \chi$.
As a consequence, the entries of this $2\times2$ matrix can be written in terms of the first two derivatives of the potential function $V'_k(\rho)$ and $V''_k(\rho)$ in terms of $\rho = \cs^{-\rk}\bar{\chi}\chi$.
We have already argued in earlier work on the mean-field approximation of group field theories~\cite{Pithis:2018bw} that this is a meaningful approximation close to critical regimes where the correlation length is expected to diverge. 
    
Field projection and the specific cyclic-melonic potential approximation allow us to explicitly derive the full non-autonomous FRG equation. 
Using the common optimized regulator function~\cite{Litim2002}
\begin{equation}
\reg = \wfr \left(k^\ks - \cs^{-\ks}C \right) \theta\left(k^\ks - \cs^{-\ks}C \right)
\end{equation}
where $\theta$ is the Heaviside function, 
we can perform the trace in the momentum space of $\textrm{U}(1)$ representations and find for the effective potential $U_k(\rho)=\mu_k\rho+V_k(\rho)$~\cite{fullpotential}
\begin{widetext}
\begin{align}
{k \partial_k U_k(\rho) } 
= {k^\ks \wfr}\left(1-\frac{\eta_k}{2} \right)  \bigg(
\frac{1}{k^\ks \wfr+ U_k'(\rho) + 2 {\rho}\, U_k''(\rho)} 
+\frac{\cct + \ccu \rk\vol{1}\ak}{k^\ks \wfr+ U_k'(\rho)}
&+\sum_{s=2}^\rk  \binom{\rk}{s}\frac{\ccu\vol{s}\ak^s}{k^\ks \wfr+ \mu_k + \frac{\rk-s}{\rk}V_k'(\rho)}\bigg) \nonumber\\
+ {k^\ks \wfr} \frac{\eta_k}{2} \bigg(
\frac{\ccu \rk \frac{\vol{1}}{3}\ak}{k^\ks \wfr + U_k'(\rho)} &+
\sum_{s=2}^{\rk} \binom{\rk}{s}\frac{\ccu s\, \frac{\vol{s}}{s+2n}\ak^s}{k^\ks \wfr + \mu_k + \frac{\rk-s}{\rk}V_k'(\rho)}\bigg)
\label{eq:boxtrace2}
\end{align}
\end{widetext}
where the particular threshold functions $\vol{s}\ak^s$ in the dimensionless variable $\ak= \cs k$ result from an integral approximation of the $s$-dimensional traces with quadratic cutoff%
\footnote{We have also explicitly calculated the trace as a sum with simplex and box cutoffs taking into account subleading orders in $\ak$ leading to the same qualitative results~\cite{fullpotential}.},
thus including the usual unit volume $\vol{s}=\pi^{s/2}/\Gamma(1+s/2)$.
The dependence on $\wfr$ is expressed in terms of the anomalous dimension $\eta_k \equiv -{k\partial_k \log\wfr}$.
As expected for a compact domain~\cite{Benedetti}, the FRG equation is a non-autonomous ordinary differential equation in the scale $k$. 
In particular there are contributions of order $\ak^s$ for $s=0,1,...,\rk$ which are due to contributions of zero modes in $\rk-s$ entries of the tensor field, respectively.

The equations resemble the FRG equation of an $\textrm{O}(2)$ model and we can specify this relation upon expanding in the average field and rescaling.
We obtain dimensionless couplings setting $\mu_k=\wfr k^2 \mr$ and for $n\ge2$
\begin{equation}\label{eq:rescaling}
\cn{n} = \wfr^{n} k^{\efd+(2-\efd)n} \cs^{(1-n)\efd} \cnr{n}
\end{equation}
where a rescaling in the volume scale $\cs$ of the group is necessary since the scaling dimension, parametrized with hindsight by an effective dimension $\efd$, is not the same as the canonical dimension $[\cn{n}]=\ks n$ in TGFT~\cite{TGFTnoncompact}.
Then, expanding the FRG equation around $\rho = 0$ yields flow equations for the dimensionless couplings
\begin{widetext}
\begin{eqnarray}\label{eq:betafunctions}
 k\partial_k \cnr{n} =  -\efd\cnr{n} + n(\efd - \ks  + \eta_k )\cnr{n} 
+\frac{1-\frac{\eta_k}{2}}{\ak^{\efd}}  {\bcb{n}(\mr,\cnr{i})}
+{\ccu} \sum_{l=1}^n   \frac{{F}_\rk^l(\ak) - \frac{\eta_k}{2} G_\rk^l(\ak)}{\ak^{\efd}}  {\bco{n}_l(\mr,\cnr{i})}
\end{eqnarray}
\end{widetext}
where $\bco{n}_l$ is the part of order $l$ in the couplings $\cnr{i}$, $i=2,...,n+1$, of the series coefficients $\bco{n}$ of the function $1+\mr+V'_k(\rho)$ around $\rho=0$ at each order $n$.
The $U''_k$-part of the full equations yields coefficients with a factor $2i-1$ for each $\cnr{i}$, that is $\bcb{n}(\mu_k, \cnr{i}):=\bco{n}(\mu_k,(2i-1)\cnr{i})$.
Furthermore, the non-autonomous structure results in polynomials
\begin{equation}
F_\rk^l(\ak) = \frac{1}{\ccu} + 2\rk \ak
+ \frac{1}{\rk^l} \sum_{s=2}^{\rk}  \binom{\rk}{s}(\rk-s)^l \vol{s}\ak^s \, , 
\end{equation}\label{eq:F}
\begin{equation}
G_\rk^l(\ak) = F_\rk^l(\ak) - \frac2{3}\rk\ak - \frac{1}{\rk^l} \sum_{s=2}^{\rk}  \binom{\rk}{s}(\rk-s)^l \frac{s\,\vol{s}}{s+2} \ak^s\,. 
\end{equation}
Except for this additional dependence on $\ak = a k$, this tower of flow equations, Eq.~\eqref{eq:betafunctions}, is exactly the one of the $\textrm{O}(2)$-invariant scalar field theory in $\efd$ dimensions. 
This is to be expected since the tensor field is complex and the action has a global $\textrm{U}(1)\simeq \textrm{O}(2)$ symmetry.
In the tensorial theory, though, the dimension $\efd$ occurs merely as a parameter which depends on the scale regime. Meaning, it can be used to obtain approximately autonomous equations in a given regime. 

Another major difference to the $\textrm{O}(2)$ model is the anomalous dimension.
The flow equation of the anomalous dimension derives from the quadratic order of field expansion of the FRG equations Eq.~\eqref{eq:Wetterich} before constant-field projection.
Since there are propagating internal momenta at one loop in TGFT one expects a significant contribution by the anomalous dimension.
Using the same threshold functions as above, we find~\cite{fullpotential}
\begin{widetext}
\begin{equation}\label{eq:etaflow}
\eta_k
= -\cnr{2}\frac{ 2(\rk-1)\ak + {\sum_{s=1}^{\rk-1} \binom{\rk-1}{s} s\,\vol{s}\ak^{s}} }
{ {\rk \ak^\efd}(1+ \mr)^2 
- {\cnr{2}} \left( \rk +
2(\rk-1)\ak  + {\sum_{s=1}^{\rk-2} \binom{\rk-1}{s} \frac{s + {\ks} }{\ks}\vol{s}\ak^{s}}
+\vol{\rk-1}\ak^{\rk-1}
\right)} \, .
\end{equation}
\end{widetext}

In the following, we will analyse the flow equations in the asymptotic regimes of large and small $\ak$ and probe the intermediate regime numerically.

\renewcommand{\ccu}{N}
\section{large $\ak$: equivalence to $\textrm{O}(\ccu)$ models}

The large-$\ak$ limit is physically interesting from two perspectives.
On a compact group $G$ with fixed volume scale $\cs$ it is equivalent to the UV limit since $\ak = a k$.
Complementary, it is also the large-volume limit 
corresponding to the theory on $\mathbb{R}^\rk$ with thermodynamic limit removing the IR regularization~\cite{TGFTnoncompact}. 
Indeed, sending $\cs\rightarrow\infty$ first, one obtains the same autonomous equations.

For rescaling in the large-$\ak$ regime one has to set $\efd = \rk-1$.
This agrees with the scaling dimension obtained from renormalization analysis~\cite{TGFTnoncompact}, in parti\-cular with scalings obtained from 
perturbative and exact renormalization group studies for field theories with tensorial interactions~\cite{TFT,KraTor2015}.
With this rescaling (and another one $\cnr{n}\mapsto \cns{n}=(2\vol{\rk-1})^{1-n}\cnr{n}/\rk$ absorbing the volume factor) we find the large-$\ak$ flow equations:
\begin{align}
k\partial_k\ms = \buv{1}(\ms,\cns{i})
=& \left(- \ks + \eta_k \right) \ms \label{eq:massflowUV}\\
&+ \rk \left(1 -\frac{\eta_k}{\efd+\ks} \right) \frac{-\cns{2}}{(1+\ms)^2} \nonumber\\
k\partial_k \cns{n} = \buv{n}(\ms,\cns{2})
=& \left(- \efd + (\efd - \ks  + \eta_k )n\right) \cns{n} \nonumber\\
&  + \left(1 -\frac{\eta_k}{\efd+\ks} \right) \bco{n}(\ms,\cns{i}) \, . \label{eq:UVequations}
\end{align}
In this regime the expansion coefficient $\bcb{n}$ does not occur. That is, only a first-derivative potential term $V'_k$ in the full expansion contributes, like in the large-$\ccu$ limit of $\textrm{O}(\ccu)$ scalar field theory (where $\ccu$ should not be confused with $\ak$). Indeed, one can check that the flow equations \eqref{eq:UVequations} are exactly the same as in that case~\cite{WetterichBergesTetradis,Delamotte} up to the factor $\rk$ in the $\ms$ equation.
We find that this factor does not change the phase structure of the theory but modifies only the exact values of scaling exponents.

Thus we have the striking result that the UV limit, or respectively the large-volume limit, of TGFT on $\textrm{U}(1)^\rk$ in the cyclic-melonic potential approximation corresponds to $\textrm{O}(\ccu)$ scalar field theory in $\efd = \rk -1$ dimensions in the large-$\ccu$ limit. 

Again, this equivalence holds only up to the anomalous dimension expected to be relatively large in TGFT.
Its flow equation for large $\ak$ is
\begin{equation}\label{eq:UVeta}
\eta_k = - \frac{(\rk-1) \cns{2}}{\ks(1+\ms)^2 - \cns{2} } \, .
\end{equation}
The crucial question is therefore whether the well-known results of the large-$\ccu$ limit in $\textrm{O}(\ccu)$ theories directly transfer to the large-$\ak$ limit in TGFT or how the specific $\eta_k$ here alters the result.

The answer is: 
We find a potential non-Gaussian fixed point not only for rank $\rk = \efd + 1 = 4$ but also for the critical case $\rk = 5$.
To this end, we solve the system of algebraic equations \eqref{eq:massflowUV}--\eqref{eq:UVeta} for couplings up to a maximal order $n=\nmax$ with vanishing derivative on the left-hand side. 
For $\rk>5$ we do not find a fixed point which converges with increasing $\nmax$.
For $\rk = 5$ there are indications for convergence to a fixed point at $(\tilde{\mu}_k,\cnr{2},...)\approx(-0.95, 0.01,...)$.
As expected, the anomalous dimension is large at this point with $\eta_k\approx-1.4$.
The scaling exponents $\theta_i$ (eigenvalues of the stability matrix $-\partial \buv{n}/\partial{\cns{m}}$, see Table~\ref{tab:exponents})
agree with the $\textrm{O}(\ccu)$ model result $\theta_i = 2i-3$~\cite{WipfBook} in sign, i.e.~all but one directions are irrelevant (have negative exponent).
In this sense it is a non-Gaussian fixed point of the same type as the Wilson-Fisher fixed point in $O(\ccu)$ models.

\begin{table*}[ht]
\caption{\label{tab:exponents}%
Scaling exponents at the non-Gaussian fixed point in the large-$\ak$ regime for $\rk=\efd+1=4$ in $(\gfb\gf)^{\nmax}$ truncation. 
}
\begin{ruledtabular}
\begin{tabular}{rcccccccccccc}
$\nmax$ & $\theta_1$ & $\theta_2$& $\theta_3$ & $\theta_4$ & $\theta_5$ & $\theta_6$ & $\theta_7$ & $\theta_8$ & $\theta_9$ & $\theta_{10}$ 
\\ 
\colrule
 6 & 0.71133 & -10.992 & -37.387 & -62.663 & -85.886 & -141.10 & \text{} & \text{} & \text{} & \text{} \\
 7 & 0.62816 & -10.250 & -35.566 & -67.696 & -86.591 & -130.58 & -201.46 & \text{} & \text{} & \text{} \\
 8 & 0.55378 & -9.4521 & -33.178 & -67.216 & -90.691 & -123.76 & -182.94 & -268.26 & \text{} & \text{} \\
 9 & 0.48821 & -8.6312 & -30.569 & -63.833 & -94.369 & -118.62 & -170.16 & -240.43 & -339.22 & \text{} \\
 10 & 0.43094 & -7.8041 & -27.876 & -59.373 & -94.466 & -115.74 & -159.03 & -220.89 & -300.97 & -411.98 \\
11 & 0.38153 & -6.9837 & -25.171 & -54.478 & -90.459 & -114.78 & -148.64 & -204.00 & -273.70 & -362.46 \\ 
\end{tabular}
\end{ruledtabular}
\end{table*}

This result implies that in the large-volume limit there might not only be be the usual phase transition with mean-field exponents~\cite{Pelissetto} above the critical case $\rk>5$ but also one captured by a non-Gaussian fixed point for $\rk=5$.
To find the exact scaling exponents it would be necessary to extend our results Tab.~\ref{tab:exponents} to higher truncations. 
At finite truncation we find the same fixed point also for $\rk=4$ but its exponents converge even slower if at all \cite{fullpotential}.

However, the large-$\ak$ equations are only valid for large $k$ on a compact group of finite volume.
Therefore, in the following we check whether this large-$k$ phase structure 
survives under the flow down to small $k$.

\section{Vanishing dimension at small scales and symmetry restoration}

In the small-$k$ limit the TGFT becomes equivalent to the $\textrm{O}(2)$ scalar theory on Euclidean space with vanishing dimension.
That is, it is necessary to rescale with effective dimension $\efd = 0$ to obtain the autonomous flow equations
\begin{equation}
k\partial_k \cnr{n} =  - \ks n \cnr{n} 
 + \bcb{n}(\cnr{i}) + \bco{n}(\cnr{i}) \, .
\end{equation}
In particular, the equation for the anomalous dimension, Eq.~\eqref{eq:etaflow} trivializes to $\eta_k = 0$.
Notably, the same result occurs for a complex scalar field on the sphere~\cite{Benedetti,Serreau}. 
In general, 
in two or less dimensions spontaneous breaking of continuous symmetries is disallowed~\cite{MerminWagner}. Therefore we find the same behaviour, in particular no phase transition for the present non-local system on $\textrm{U}(1)^{r}\cong (S^1)^{\times r}$ at small $\ak$. 
This is also nourished by previous works using mean-field arguments~\cite{Pithis:2018bw} as well as FRG studies applied to TGFT~\cite{BGBeOr2015,TGFTnoncompact,TGFT1,TGFT2} which, however, have remained inconclusive on this matter so far. 

\begin{figure}[ht]
\includegraphics[scale=.35]{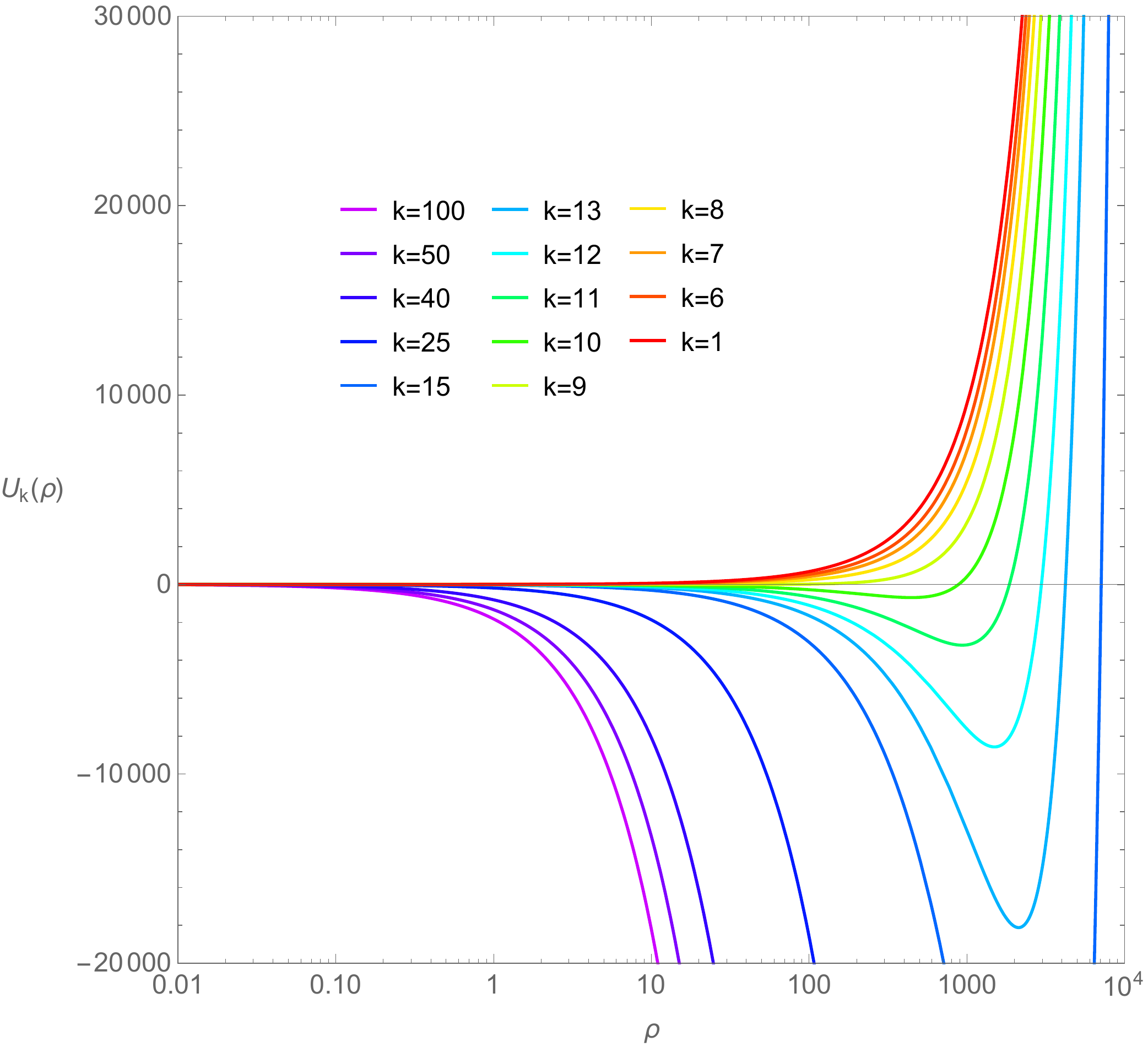}
\centering
\caption{Flow of the dimensionful potential $U_k = \mu_k\rho + V_k(\rho)$ in the $\nmax=4$ truncation for rank $\rk=5$ between $k=100$ and $k=1$ with $a=1$ for initial conditions at $\Lambda=100$ close to the UV non-Gaussian fixed point in that truncation:
$Z(\Lambda)=1$, $\mu(\Lambda)=-0.86 \Lambda^2$, $\lambda_2(\Lambda)=0.090\Lambda^0$, $\lambda_3(\Lambda)=0.084\Lambda^{-2}$ and $\lambda_4(\Lambda)=0.075\Lambda^{-4}$. 
}\label{figure:flowofpotentialdimfulm=4}
\end{figure}

\begin{figure}[ht]
\centering
 \includegraphics[scale=.35]{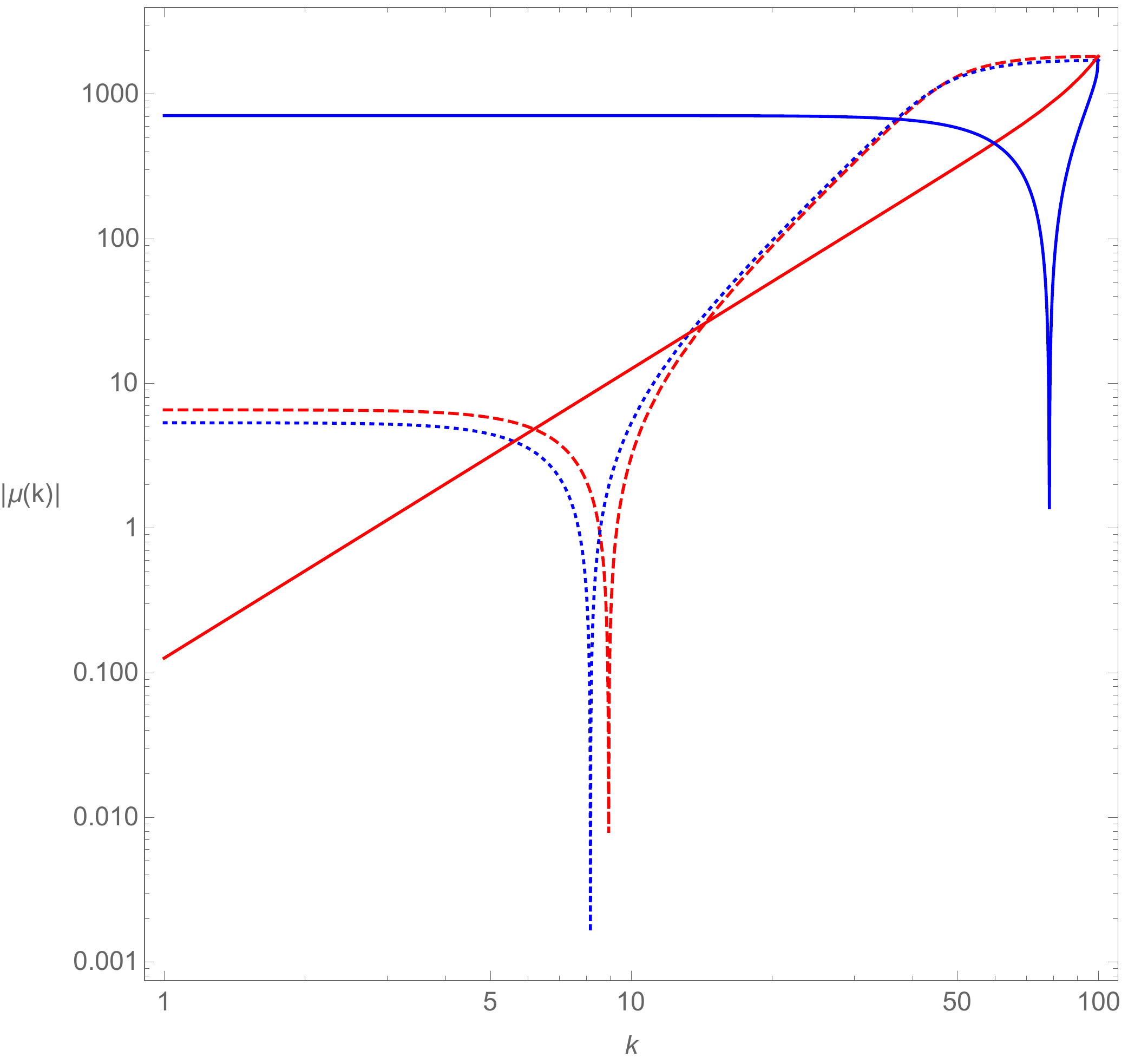}
\caption{Flow of the modulus of $\mu(k)$ for $\rk=5$ in the $\nmax=4$ truncation for: (I) the system of non-autonomous $\beta$-functions~\eqref{eq:betafunctions} with $\cs=1$ (dashed lines) and (II) the set of autonomous $\beta$-functions in the large-volume limit~\eqref{eq:UVequations} (continuous lines). Initial conditions are the same as in Fig.~\ref{figure:flowofpotentialdimfulm=4} except for $\mu(\Lambda)= -0.91\Lambda^2$ (red) and $\mu(\Lambda)=-0.85 \Lambda^2$ (blue).
}\label{figure:flowofmassandminumum}
\end{figure}

In fact, the non-Gaussian fixed point in the UV vanishes quickly under the flow as we confirm with numerical calculations. 
To this aim, we solve the full non-autonomous equations, Eq.~\eqref{eq:betafunctions}, for the dimensionful potential $U_k(\rho)=\mu_k\rho + V_k(\rho)$ from $k=\Lambda$ to small $k$.
Starting with any potential which explicitly exhibits spontaneous breaking of the global $\textrm{U}(1)$ symmetry in the UV, i.e. $\rho_\Lambda\ne0$, we observe symmetry restoration at some finite value of $k$. We exemplify this point for the concrete case of a rank-$5$ TGFT, however, the same holds true for any rank $\rk$.
The flow of the dimensionful potential in the $\nmax=4$ truncation is reported in Fig.~\ref{figure:flowofpotentialdimfulm=4}; flows of $\mu_k$ in both phases in the large-volume limit are contrasted with the flow in the compact case with same initial conditions in Fig.~\ref{figure:flowofmassandminumum}. 

Let us emphasize that the underlying reason for symmetry restoration is very universal and extends far beyond our melonic truncation.
The source for symmetry restoration are the zero modes in the spectrum of the theory.
It is the $\rk$-fold zero mode which leads to the constant term in the non-autonomous part $F_\rk^l$ in the flow equations, Eq.~\eqref{eq:betafunctions}. 
This lowest-order term dominates in the limit $k\rightarrow 0$ and demands a scaling of the couplings with dimension $\efd=0$. 
This mechanism is independent of the combinatorial structure of the interactions.
It relies simply on the compact domain of the fields
which is in agreement with the general lore that there are no true phase transitions in systems of finite size~\cite{JZJ,Strocchi}. 
Though we have illustrated the phenomenon here with the example of a cyclic-melonic potential approximation at arbitrary order, we therefore expect symmetry restoration for the full potential of any TGFT with discrete spectra containing isolated zero modes.

\section{Conclusions}

Our main result is that there are no phase transitions between a broken and unbroken phase of the global $\textrm{U}(1)$ symmetry in TGFT on $\textrm{U}(1)^\rk$ 
at any rank $\rk$ in the cyclic-melonic potential approximation.
Thereby, we have performed for the first time a local-potential approximation at any order and on any scale in TGFT.
We have argued that this result essentially relies on the isolated zero mode in the spectrum.
On this ground we conjecture that this result extends to field theories with any tensor-invariant interactions, on any compact group $G$.
In the large-volume limit corresponding to the theory on $\mathbb{R}^\rk$ there is a correspondence to $\rk-1$ dimensional $\textrm{O}(\ccu)$ models in the large-$\ak$ limit and due to the tensor-specific flow of the anomalous dimension there is a possible non-Gaussian fixed point even in the critical case rank $\rk=5$.
One might take this equivalence as a hint for a new type of examples of holographic duality~\cite{SeSu2002,KlPo2002} since rank-$\rk$ TGFT is related to gravitational theories in $\rk$ spacetime dimensions.

More realistic models of quantum gravity should therefore be defined on non-compact groups to open the room for a phase transition to continuum spacetime.
It remains to be checked how our result of a universal symmetry restoration transfers to group field theories characterized by a closure constraint~\cite{GFT},
models with an additional gravitational (Holst-Plebanski) constraint~\cite{Perez}, as well as models with an additional matter reference frame~\cite{Matterdof} reminiscent of SYK models~\cite{SYKreview}.
But one would assume that the zero-mode effect persists in such models as well.
Only if the tensorial degrees of freedom are not dynamic as in actual SYK-type models~\cite{SYKtype} the zero-mode effect would be absent. 
In any case, if there are effectively some non-compact degrees of freedom as for example reference frame fields~\cite{Matterdof} these should contribute non-vanishing dimensions.
It is therefore of interest to study the FRG of such models.
Still, in models with dynamical tensorial degrees of freedom which are the ones related to quantum gravity, the obvious way out is to consider models on non-compact groups.
For a physical theory of gravity it might be necessary anyway,
as holonomies in gravity are captured by the Lorentzian group which, in particular, covers the causal structure of spacetime.  

\subsection*{Acknowledgments}
The authors thank D. Benedetti, J. Ben Geloun, A. Duarte Pereira, A. Eichhorn, D. Oriti and R. Percacci for discussions and critical remarks on an earlier version of this text, and in particular S. Carrozza for the inspiration to study cyclic-melonic interactions for a local-potential approximation.

The work of AGAP leading to this publication was supported by the PRIME programme of the German Academic Exchange Service (DAAD) with funds from the German Federal Ministry of Education and Research (BMBF).
The work of JT was funded by the Deutsche Forschungsgemeinschaft (DFG, German Research Foundation) in two ways,
primarily under the author's project number 418838388 and
furthermore under Germany's Excellence Strategy EXC 2044 –390685587, Mathematics M\"unster: Dynamics–Geometry–Structure.

\end{document}